\documentclass[intlimits,twoside,a4paper]{article}
\usepackage{amsmath,amssymb}
\usepackage{graphicx}
\usepackage{wrapfig}
\usepackage[T2A]{fontenc}
\usepackage[cp1251]{inputenc}

\usepackage[eqsecnum]{cmpj}

\issue{2011}{14}{3}{33601}

\doinumber{10.5488/CMP.14.33601}

\title[Adsorption of hard spheres]%
{Adsorption of hard spheres: structure and effective density according to the potential distribution theorem}
\author[L.L.~Lee, G.~Pellicane]{L.L. Lee\refaddr{label1}, G.~Pellicane\refaddr{label2}}
\addresses{
\addr{label1} Department of Chemical \& Materials Engineering, California State University, \\Pomona, California, USA
\addr{label2} School of Physics, University of Kwazulu-Natal, Private Bag X01 Scottsville, \\
3209 Pietermaritzburg, South Africa
}
\authorcopyright{L.L.~Lee, G.~Pellicane, 2011}

\date{Received April 11, 2011}

\begin{document}

\maketitle

\begin{abstract}
We propose a new type of effective densities via the potential distribution theorem.  These densities are for the sake of enabling the mapping of the free energy of a \emph{uniform} fluid onto that of a \emph{nonuniform} fluid.  The potential distribution theorem gives the work required to insert a test particle into the bath molecules under the action of the external (wall) potential.  This insertion work $W_{\mathrm{ins}}$ can be obtained from Monte Carlo (MC) simulation (e.g. from Widom's test particle technique) or from an analytical theory.  The pseudo-densities are constructed thusly so that when their values are substituted into a \emph{uniform-fluid} equation of state (e.g. the Carnahan-Starling equation for the hard-sphere chemical potentials), the MC nonuniform insertion work is reproduced. We characterize the pseudo-density behavior for the hard spheres/hard wall system at moderate to high densities (from $\rho^{\ast} = 0.5745$ to $0.9135$).  We adopt the MC data of Groot et al. for this purpose. The pseudo-densities show oscillatory behavior out of phase (opposite) to that of the singlet densities.  We also construct a new closure-based density functional theory (the \emph{star-function based density functional theory}) that can give accurate description of the MC density profiles and insertion works.  A viable theory is established for several cases in hard sphere adsorption.
\pacs 68.43.De
\end{abstract}

\section{Introduction}

Inhomogeneous fluids pose, due to their vast varieties, challenges to the liquid-state theories, and are of capital importance in many industrial processes. The nonuniformity in the fluid can be produced by a one-body potential: such as the surface potential from the interfaces separating two or three coexisting phases (solid/liquid/vapor phase transitions), electrical fields acting on ionic species (the electric double layer), or magnetic fields on ferromagnetic liquids. The conventional classical density functional theory (DFT) employed to deal with such systems in the last half century has been based on the Hohenberg-Kohn~\cite{hohenberg} formulation of a grand potential, $\Omega$, that is expressed in terms of an intrinsic Helmholtz free energy functional $F[\rho_w^{(1)}]$ (IHFEF):
\begin{equation}
\label{eq1}
\Omega = F[\rho_w^{(1)}] + \int  \rd \vec{r} \rho_w^{(1)}(\vec{r};[w]) \lbrace w(\vec{r})-\mu_0 \rbrace
\end{equation}
where $\rho_w^{(1)}$  is the singlet (nonuniform) density, $w(r)$ the external (1-body) potential, and $\mu_0$ the chemical potential of the bulk fluid.  At equilibrium, the grand potential is minimized with respect to the singlet $\rho_w^{(1)}$, and it yields the Euler-Lagrange equation
\begin{equation}
\label{eq2}
\frac{\delta\beta\Omega}{\delta\rho_w^{(1)}(\vec{r})} = \frac{\delta\beta F[\rho_w^{(1)}]}{\delta\rho_w^{(1)}(\vec{r})} -\beta\mu_0 + \beta w(\vec{r})=0
\end{equation}
where $\beta$ = $1/(kT)$: $k$ is the  Boltzmann constant, $T$ is the absolute temperature.
	The IHFEF can be separated into two parts: the ideal part $F_{\mathrm{id}}$ (which is known: $F^{\mathrm{id}}[\rho] = \int  \rd \vec{r} \rho_w^{(1)}(\vec{r}) [ \ln \rho_w^{(1)}(\vec{r})\Lambda^3 - 1]$) and the excess part $F^{\mathrm{ex}}$. Thus $F[\rho_w^{(1)}]$ = $F^{\mathrm{id}} + F^{\mathrm{ex}}$. Recently, we have proposed a star-function based density functional theory~\cite{lee1,lee2} (s-DFT) which expresses $F^{\mathrm{ex}}$ succinctly as
\begin{equation}
\label{eq3}
-\beta F_{[ \rho ]}^{\mathrm{ex}} = -\beta F_{0}^{\mathrm{ex}} + \int \rd \vec{r} C_{0}^{(1)}(\vec{r})\delta \rho_w(\vec{r}) + \frac{1}{2} \int \rd \vec{r} \rd \vec{r}\,' C_{0}^{(2)}(\vec{r},\vec{r}\,')\delta \rho_w(\vec{r}) \delta \rho_w(\vec{r}\,') + S_w^{\ast}
\end{equation}
where $S_w^{\ast}$ is the star function~\cite{lee3} defined as the ``primitive'' of the bridge function $B_w$
\begin{equation}
\label{eq4}
S_w^{\ast} = \int \rd \vec{r} \frac{\delta \rho_w(\vec{r})}{\gamma_w(\vec{r})}  \int_{\gamma_0}^{\gamma_1} \rd x B_w(x)
\end{equation}
and $C_{0}^{(1)}(\vec{r})$ is the singlet direction correlation function ($1-DCF$), $C_{0}^{(2)}(\vec{r},\vec{r}\,')$ is the pair direct correlation function ($2-DCF$). $F_0^{\mathrm{ex}}$ is the bulk fluid free energy.  Subscripts $0$ and $w$ indicate uniform and nonuniform fluid properties, respectively. $\delta \rho_w$ = $(\rho_w^{(1)} - \rho_0)$ is the difference between the nonuniform density $\rho_w^{(1)}$ and the uniform density $\rho_0$ and $\gamma_w$ is the indirect correlation. We note that if the bridge function $B_w$ can be obtained exactly, the star function $S_w^{\ast}$ will consequently be exact from equation~(\ref{eq4}), thus the excess free energy $F^{\mathrm{ex}}$ in equation~(\ref{eq3}) will also be exact. The Euler-Lagrange (EL) equation~(\ref{eq2}) can be expressed as
\begin{equation}
\label{eq5}
\rho_w(\vec{r}) = \rho \exp\left[-w(\vec{r}) + C_{w}^{(1)}(\vec{r}) - C_{0}^{(1)}(\vec{r})\right]
\end{equation}
where $C_{w}^{(1)}$ is the nonuniform singlet direct correlation function ($1-DCF$).  Alternatively, it is possible to show~\cite{lee1} via functional expansions that equation~(\ref{eq5}) can be written as
\begin{equation}
\label{eq6}
\rho_w(\vec{r}) = \rho \exp\left[-w(\vec{r}) + \gamma_w(\vec{r}) + B_w(\vec{r})\right].
\end{equation}
Thus, we have the equality~\cite{lee2}
\begin{equation}
\label{eq7}
C_{w}^{(1)}(\vec{r}) - C_{0}^{(1)}(\vec{r}) = \gamma_w(\vec{r}) + B_w(\vec{r}).
\end{equation}
Equations~(\ref{eq3}) and (\ref{eq6}) constitute the basis of the s-DFT theory.

In this article, we shall explore two essential aspects of nonuniform fluids:
(1) the effective density that enables the mapping between the uniform-fluid
free energies and the nonuniform-fluid free energies; (2) the closure relation
between the bridge function and other correlation functions. The first task is
mediated through the potential distribution theorems~\cite{widom,henderson};
the second is based on a successful uniform liquid theory~\cite{lee4} developed in recent years.
Historically, there have been interchanges and cross-fertilization of uniform
and nonuniform liquid theories~\cite{denton,brenan}.  In this paper we test the new formulations
on hard spheres adsorbed on hard planar walls.

In weighted-density density functional approaches (WDA)~\cite{evans,henderson2} the ``mapping'' of
the free energy is done through some \emph{weighted density}: $\overline{\rho}$. The weighted density is obtained from the true density $\rho_w^{(1)}$  via a convolution integral with a \emph{weighting function}, $\omega$, as the kernel: i.e. $\overline{\rho}(\vec{r}) = \int \rd\vec{r}\,' \rho_w^{(1)}(\vec{r} \,')\omega(\vec{r},\vec{r}\,')$. We shall, however, develop an effective density without going through this weighting procedure.  This is made possible by the potential distribution theorem (PDT), since PDT naturally lends itself to yield the work of insertion (the free energy of an inhomogeneous system).

The PDT is a fundamental theory of statistical mechanics~\cite{widom,henderson,henderson2,beck,chipot}.  It is concerned with the work required to insert test particles in an equilibrium ensemble representing a uniform fluid or a nonuniform fluid.  In an earlier paper~\cite{lee5} we have generalized the PDT from the chemical potential, to the cavity functions, and to higher-order correlation functions.  The PDT has previously been extended to nonuniform systems~\cite{widom,henderson}.  However, here we distinguish two types of test particles. The first type is for a test particle subject to the wall force as well as to the forces of the bath molecules. This is the commonly studied type.  The second type is for a test particle \emph{free} of the wall force. Its interaction with the wall is intentionally removed (similar to the cavity function in homogeneous fluids). The PDT of the second kind is directly related to the Euler-Lagrange equation~\cite{lee2}. The PDT establishes a connection among (i) the insertion work $W_{\mathrm{ins}}$, (ii) the bulk fluid chemical potential $\mu_0$, and (iii) the singlet distribution function  $\rho_w^{(1)}$. To illustrate the PDT, we carry out calculations on the system of hard spheres on a hard wall (HS/HW), for which MC data are available (e.g. from Groot et al.~\cite{groot}).

The mapping between the uniform and the nonuniform systems can be constructed as follows: given an equation of state for the uniform fluid (which is utilized to calculate the uniform chemical potential $\beta\mu_0$), we solve for the hypothetical densities $\rho_{\mathrm{pseudo}}(z)$ that will reproduce the insertion works  $W_{\mathrm{ins}}(z)$ obtained from the MC, i.e. $\beta\mu_0(\rho_{\mathrm{pseudo}}(z)) = \beta W_{\mathrm{ins}}(z)$.  This density profile is an artificial \emph{construct} and is called the \emph{pseudo-density} $\rho_{\mathrm{pseudo}}(z)$.  Its sole role is to reproduce the nonuniform free energy via the bulk equation of state.  For instance, in the HS/HW system, the bulk fluid equation is taken to be the Carnahan-Starling (CS)~\cite{carnahan} equation which is known to be highly accurate.  It yields the uniform hard-sphere chemical potential $\beta\mu_0$. If we know the insertion work  $\beta W_{\mathrm{ins}}(z)$ (from MC) in HS/HW, a series of values of $\rho_{\mathrm{pseudo}}(z)$ can be generated so that the equality   $\beta\mu_0^{\mathrm{HS}} = \beta W_{\mathrm{ins}}(z)$ holds.  The $\rho_{\mathrm{pseudo}}(z)$ obtained thusly will enable the mapping of the free energies (  $\beta\mu_0^{\mathrm{HS}}$) of the uniform fluid to those ($\beta W_{\mathrm{ins}}$) of the nonuniform fluid.  In this respect, the pseudo-densities perform precisely the same function as the weighted densities $\overline{\rho}(z)$ in the WDA. The difference is that our approach is based on the PDT, an exact theory, not on approximate theories (such as the Percus-Yevick equation~\cite{percus}).

Should the bridge function $B_w$ be known, one can obtain the fluid structure at the wall via the Euler-Lagrange equation~(\ref{eq6})~\cite{lee1,lee2}. $B_w$ as a correlation function is well defined in terms of infinite series of cluster diagrams~\cite{hiroike} or functional expansions~\cite{iyetomi}. However, direct evaluation from its definition is numerically intractable.  But it can be inverted from machine data (as in \emph{reverse engineering}~\cite{duh}) if we consider equation~(\ref{eq6}) as the defining equation~\cite{iyetomi}.  The inverted $B_w$ can be given as a set of numerical data~\cite{lee2}, say, in the form of tabulated entries; or it can be expressed as a function $B_w = f(r)$ for some analytical function $f$ (e.g. a polynomial). Another approach, and a more appealing one is to formulate a liquid theory ($\grave{a}$ la mode of the Percus-Yevick equation~\cite{percus}) that relates $B_w$ to other correlation functions $B_w = f(\gamma)$ , $\gamma$ being a well-defined quantity in liquid state theory. This in uniform liquid theory is called the \emph{closure relation}.  This relation may or may not exist, because $B_w$ is in fact a functional~\cite{lee2,lee6} $B_w = \Phi [ \rho_w^{(1)},w ]$ of the singlet density $\rho_w^{(1)}$  and/or the external potential $w$ ($\Phi [. ]$ being the functional sign). Thus, it may not be simply related to any one correlation function. For uniform liquids, the determination of the function $B = f(\gamma)$, $\gamma$ being some correlation, is already a major task (examples such as the Rogers-Young~\cite{rogers} (RY) closure or the Martynov-Sarkisov~\cite{martynov} (MS) closure).  We have proposed a theory upon resummation of the functional expansions of $B_w$ in \cite{lee2}. This resummation seems to work well for a number of pair potentials.  For nonuniform fluids, the PY and hypernetted-chain~\cite{morita} (HNC) closures have been tried and shown not to work well in the past~\cite{evans2}.  And this prompted the proposition that closure-based approaches might not be suitable for nonuniform systems.  In this work, we shall propose a zero-separation type (ZSEP) closure derived from the study of uniform liquid theory~\cite{lee4}.  This closure will be tested here on selected cases of the HS/HW systems. Once the singlet density is obtained from equation~(\ref{eq6}) given $B_w$, one can use the Euler-Lagrange equation~(\ref{eq5}) to obtain the 1-DCF $C_{w}^{(1)}$.  The 1-DCF is simply related to the insertion work as
\begin{equation}
\label{eq8}
\beta W_{\mathrm{ins}}(\vec{r}) = - C_{w}^{(1)}(\vec{r}).
\end{equation}

Thus, we can access the insertion work through~(\ref{eq5}), using the MC data~\cite{groot} on $\rho_w^{(1)}$  as input.  The  thusly obtained $\beta W_{\mathrm{ins}}$ is considered as the MC-derived insertion work. In addition, we shall construct a theoretical method for obtaining the insertion work.

Section~2 gives the potential distribution theorems for nonuniform fluids with general interaction forces.  Section~3 shows the calculations of the effective densities based on the MC-derived insertion works for the hard spheres/hard wall system. The mapping of the nonuniform free energies is mediated through an accurate equation of state by Carnahan and Starling. We characterize the behavior of $\rho_{\mathrm{pseudo}}$ and compare with the commonly used weighted densities found in literature. In section~4 we test the new closure theory (ZSEP) on the same systems. We compare the theoretical structures and free energies with the MC data.  In section~5 we draw the conclusions.

\section{Potential distribution theorems for nonuniform systems}

In this section, we shall introduce the potential distribution theorems for two types of test particles. One type of test particle is the well-known one, i.e. it is subject to interactions with the bath molecules as well as to forces of the wall. The second type is the one that interacts only with bath molecules and not with the wall.  The latter result will be the PDT that we shall use in this article.
We start with the averages in a canonical ensemble.  (This can also be done in the grand canonical ensemble~\cite{widom,henderson}). The N-body system consists of $N$ fluid molecules and they border on one side at a solid surface. The Hamiltonian $H_N$ is written as
\begin{equation}
\label{eq9}
H_N(\vec{p}^N,\vec{r}^N) = K_N\left(\vec{p}^N\right) + U_N\left(\vec{r}^N\right)+W_N\left(\vec{r}^N\right),
\end{equation}
where $\vec{r}^N \equiv (\vec{r}_1,\vec{r}_2,\dots ,\vec{r}_N)$ is a shorthand for the $N$-vector of the positions of $N$ particles, and $p_N$ is the $N$-momenta vector, $K_N$ is the kinetic energy, while
\begin{equation}
\label{eq10}
U_N(\vec{r}^N) = \sum_{1=i<j}^{N-1}\sum_{j=2}^{N}u^{(2)}(r_i,r_j)
\end{equation}
is the total potential energy between fluid particles (assumed to be pairwise additive), and
\begin{equation}
\label{eq11}
W_N(\vec{r}^N) = \sum_{k=1}^{N}w(r_k)
\end{equation}
is the sum of one-body energies arising from the external (\emph{wall}) potential $w$. This is the potential energy that \emph{drives} the inhomogeneities in the system.

Next, we consider an (N-1)-body system: the sum of pair energies becomes
\begin{equation}
\label{eq12}
U_{N-1}(\vec{r}^{\,N-1}) = \sum_{1=i<j}^{N-2}\sum_{j=2}^{N-1}u^{(2)}(r_i,r_j)
\end{equation}
and the sum of one-body energies
\begin{equation}
\label{eq13}
W_{N-1}(\vec{r}^{N-1}) = \sum_{k=1}^{N-1}w(\vec{r}_k).
\end{equation}

A test particle is introduced to the (N-1)-body system as particle with label ``N'' located at the distance $r_N$.  This test particle can interact with the other $N-1$ \emph{bath} molecules through the pair potential $u^{(2)}(r_i,r_N)$, $i < N$, as well as with the wall through the one-body potential $w(\vec{r}^N)$.  This particle is designated the \emph{type-1 test particle}.  The excess potential energy $\Psi_w(\vec{r}_N)$ due to the presence of type-1 test particle is
\begin{equation}
\label{eq14}
\Psi_w(\vec{r}_N) = U_N(\vec{r}^N) - U_{N-1}(\vec{r}^{N-1}) + W_N(\vec{r}^N) - W_{N-1}(\vec{r}^{N-1}) = \sum_{i=1}^{N-1}u^{(2)}(\vec{r}_{i},\vec{r}_{N}) + w(\vec{r}_N).
\end{equation}

On the other hand, the \emph{type-2 test particle}, by choice, does not interact with the wall ($w(\vec{r}_N) = 0$).  The excess potential energy $\Psi_{-w}(\vec{r}_N)$ is (subscript ``--w'' stands for ``without the wall interaction'')
\begin{equation}
\label{eq15}
\Psi_{-w}(\vec{r}_N) = \sum_{i=1}^{N-1}u^{(2)}(\vec{r}_{i},\vec{r}_{N}).
\end{equation}

The \emph{potential distribution theorem} for the \emph{first-type test particle} (interacting with $w$) is obtained from the (N-1)-body canonical ensemble average of $\Psi_{w}(\vec{r}_N)$
\begin{equation}
\label{eq16}
\ln \langle \exp\left[-\beta\Psi_{w}\right]\rangle_{N-1;w} = -\beta \mu_w + \ln \left[\rho_w^{(1)}(\vec{r}_N)\Lambda^3\right]
\end{equation}
since
\begin{eqnarray}
\label{eq17}
\langle \exp\left[-\beta\Psi_{w}\right]\rangle_{N-1;w} &=& \frac{1}{Q_{N-1}} \int \rd \vec{r}^{N-1} \exp\left[ -\beta U_{N-1} - \beta W_{N-1} \right] \nonumber\\
&&{}\times \exp \left\{ -\beta \left[ \sum_{i=1}^{N-1}u^{(2)}(r_{iN}) + w(\vec{r}_N) \right] \right\} .
\end{eqnarray}

Furthermore, from the definition of the singlet density $\rho_w^{(1)}(\vec{r}_N)$
\begin{equation}
\label{eq18}
\rho_w^{(1)}(\vec{r}_N) = \frac{N}{Q_{N}} \int \rd \vec{r}^{N-1} \exp\left[ -\beta U_{N} -\beta W_N \right].
\end{equation}

Equation~(\ref{eq16}) has been obtained earlier~\cite{widom,henderson}.  Note that $Q_N$ is the N-body configurational integral; $\Lambda$ is the de Broglie wavelength; and
\begin{equation}
\label{eq19}
\beta \mu_w = \ln \frac{Q_{N-1}}{Q_{N}} + \ln (\rho \Lambda^3).
\end{equation}

The \emph{potential distribution theorem} for the \emph{type-2 test particle} (without $w$) is obtained instead from the (N-1)-body ensemble average of the \emph{wall-less} excess potential $\Psi_{-w}(\vec{r}_N)$ equation~(\ref{eq15}), i.e.
\begin{equation}
\label{eq20}
\ln \langle \exp\left[-\beta\Psi_{-w}\right]\rangle_{N-1;w} = -\beta \mu_w + \beta w(\vec{r}_N) + \ln \left[\rho_w^{(1)}(\vec{r}_N)\Lambda^3\right]
\end{equation}
since
\begin{eqnarray}
\label{eq21}
\langle \exp\left[-\beta\Psi_{-w}\right]\rangle_{N-1;w} &=& \frac{\exp \left[ \beta w(\vec{r}_N) \right]}{Q_{N-1}} \int \rd \vec{r}^{N-1} \exp\left[ -\beta U_{N-1} - \beta W_{N-1} \right] \nonumber \\
&&{}\times \exp \left\{ -\beta \left[ \sum_{i=1}^{N-1}u^{(2)}(r_{iN}) + w(\vec{r}_N) \right] \right\}.
\end{eqnarray}

The insertion work $W_{\mathrm{ins}}$ required to insert a type-2 test particle with the excess potential $\Psi_{-w}(\vec{r}_N)$  is thus (noting that $\beta \mu_w = \beta \mu_0^{\mathrm{ex}} + \ln (\rho_0\Lambda^3)$)
\begin{eqnarray}
\label{eq22}
\beta W_{\mathrm{ins}} &=& -\ln \langle \exp\left[-\beta\Psi_{-w}\right]\rangle_{N-1;w} = \beta \mu_w - \beta w(\vec{r}_N) - \ln \left[\rho_w^{(1)}(\vec{r}_N)\Lambda^3\right]  \nonumber \\{}&=& \beta \mu_0^{\mathrm{ex}} - \beta w(\vec{r}_N) - \ln \left[\rho_w^{(1)}(\vec{r}_N)/\rho_0\right].
\end{eqnarray}

The quantity $W_{\mathrm{ins}}$ can be directly simulated via the MC method using the Widom particle-insertion technique~\cite{widom,frenkel}, or from the EL equation (\ref{eq5}). $W_{\mathrm{ins}}$ can also be considered as the \emph{intrinsic} work of insertion.

\section{Effective density and work of insertion}

For the hard sphere system, we adopt the MC data by Groot et al.~\cite{groot} as the basis of calculation. The number densities chosen for the hard spheres are $\rho^{\ast}$ = $0.5745$, $0.715$, $0.758$, $0.813$, and $0.9135$.  The hard spheres are adsorbed on a planar hard wall stretching over the x-y plane. Thus, the fluid-fluid pair potential $u^{(2)}$ is
\begin{equation}
\label{eq23}
\begin{cases}
u^{(2)}(r) = \infty, & r \leqslant  \sigma,\\
u^{(2)}(r) = 0, & r > \sigma.
\end{cases}
\end{equation}

The wall potential w on the fluid particles (in the z-direction perpendicular to the wall) is
\begin{equation}
\label{eq24}
\begin{cases}
w(z) = \infty, &  \forall z \leqslant  \frac{\sigma}{2}\,,\\
w(z) = 0 ,& \forall z > \frac{\sigma}{2}\,.
\end{cases}
\end{equation}

\begin{figure}[!ht]
\centerline{\includegraphics[width=8cm]{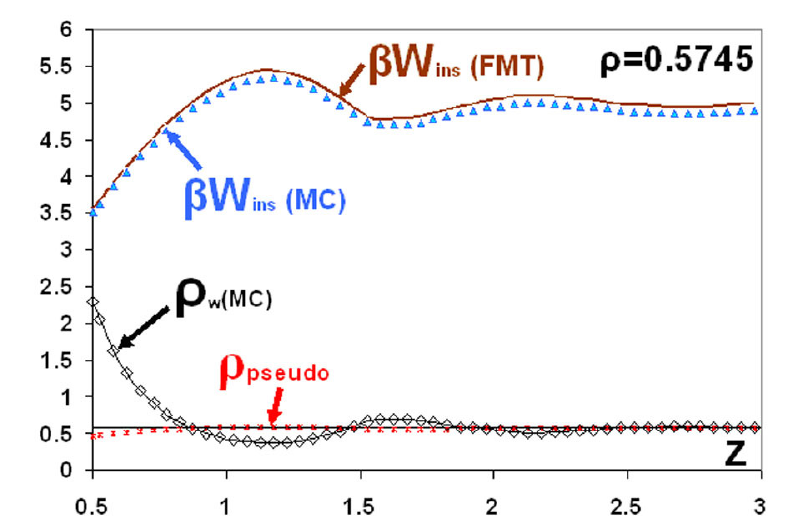}}
  \caption{The effective (pseudo) density $\rho_{\mathrm{pseudo}}(z)$ (symbol $\ast$; from PDT) and the insertion work $\beta W_{\mathrm{ins}}$ (symbol $\triangle$; from MC) at the bulk hard-sphere density $\rho^{\ast} = 0.5745$ for the HS/HW system. Symbol ($\diamondsuit$) $\equiv$ the singlet density profile  $\rho_w^{(1)}(z)$ from MC data of Groot et al.~\protect\cite{groot}.  The brown line is the insertion work $\beta W_{\mathrm{ins}}$ produced by the FMT theory.  $Z$ $\equiv$ distance from the wall. Unit of length is taken to be the hard-sphere diameter $\sigma$.  The horizontal line is at $\rho^{\ast} = 0.5745$ (a guide for the eye). The pseudo-density $\rho_{\mathrm{pseudo}}(z)$ oscillates weakly around the bulk value and in opposite periodicity as compared to $\rho_w^{(1)}(z)$.  The insertion work $\beta W_{\mathrm{ins}}$ also oscillates synchronously with $\rho_{\mathrm{pseudo}}(z)$  but oppositely to $\rho_w^{(1)}(z)$.}
  \label{fig1}
\end{figure}
\begin{figure}[ht]
\centerline{\includegraphics[width=8cm]{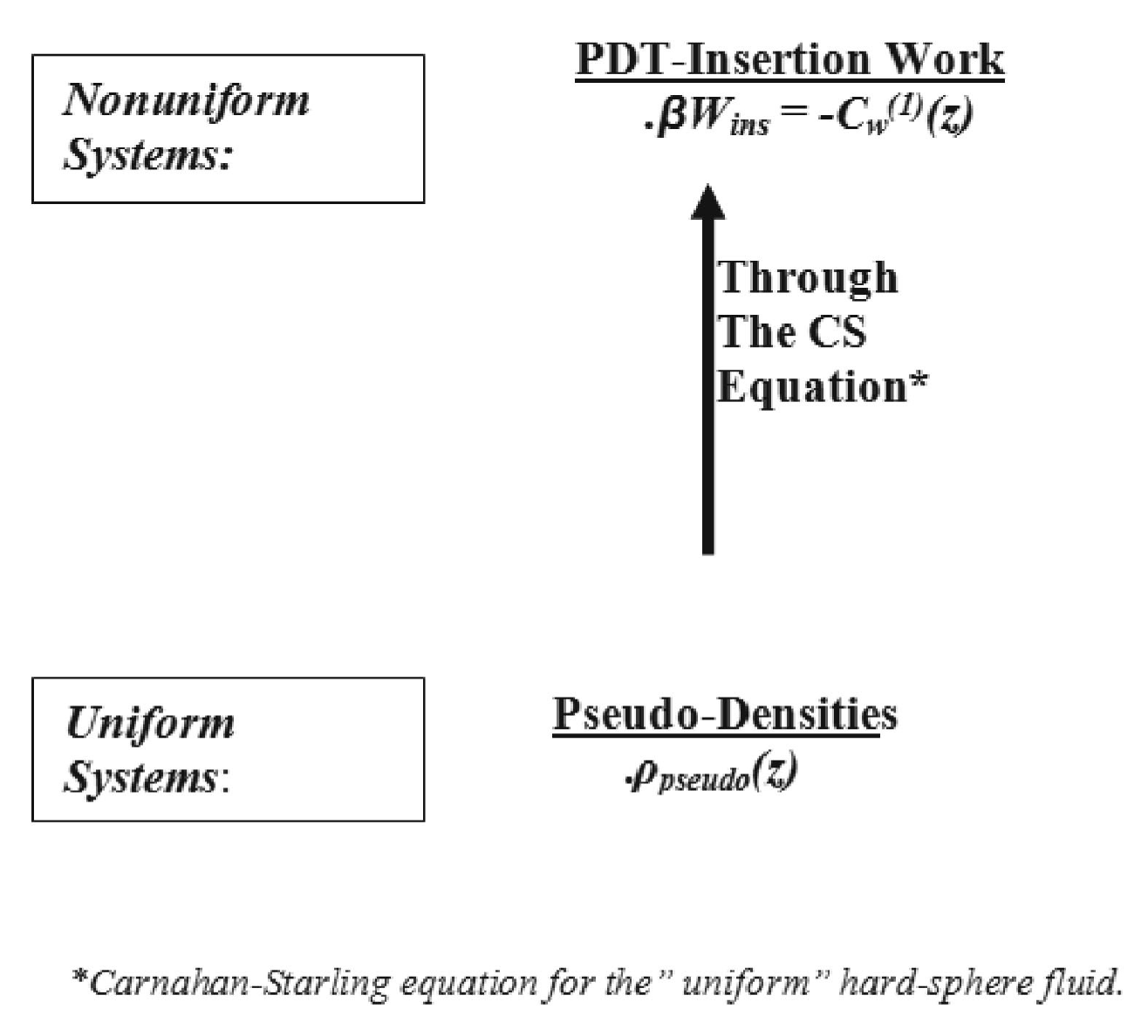}}
\caption{``Mapping'' of properties between the non-uniform system and the uniform systems.}  \label{tab1}
\end{figure}

Thus, the closest approach of the center of a hard sphere to the wall is at $z ={\sigma}/{2}$ (we shall use the hard sphere diameter $\sigma$ as unit of length below). For $\rho^{\ast} = 0.5745$, the pure hard-sphere fluid excess chemical potential   $\beta\mu_0$ is $4.897$, as calculated from the CS equation.  The singlet density profile $\rho_w^{(1)}(z)$ has been obtained by Groot et al.~\cite{groot} via MC simulation and is displayed in figure~\ref{fig1} (see the diamond symbols $\Diamond$).  The insertion work $\beta W_{\mathrm{ins}}(z)$ is identified as $-C_w^{(1)}(z)$ (equation~(\ref{eq8}) i.e. the (negative) 1-DCF, which is obtained from the MC data via the EL equation (equation~(\ref{eq5}). The insertion work is plotted as triangles ($\blacktriangle$). We remark that both the singlet density $\rho_w^{(1)}(z)$ and the insertion work $\beta W_{\mathrm{ins}}(z)$ are ``real'' quantities, i.e. they are measurable quantities for the system of interest.  We next formulate the pseudo-densities as mentioned earlier in section~1. We ask at what \emph{fictitious} densities $\rho_{\mathrm{pseudo}}(z)$ for a \emph{uniform} HS fluid (which follows the CS equation), will we obtain a chemical potential value $\beta\mu_0$  that is equal to the MC nonuniform insertion work $\beta W_{\mathrm{ins}}(z)$
\begin{equation}
\label{eq25}
\beta \mu_0 = \frac{4 \eta - 3\eta^2}{(1 - \eta)^2} + Z'^{\mathrm{HS}} = \beta W_{\mathrm{ins}}(z),
\end{equation}
where $\eta = \frac{\pi}{6}\rho_{\mathrm{pseudo}}(z)\sigma^3$. In figure~\ref{tab1}, we show this mapping via a diagram.  Note that $Z'^{\mathrm{HS}}$ is the non-ideal compressibility factor from the CS equation.  The pseudo-density $\rho_{\mathrm{pseudo}}(z)$ thus calculated is also plotted (as asterisks $\ast$) in figure~\ref{fig1}.   We observe that the pseudo-density stays fairly flat and close to the bulk value of $0.5745$, oscillating only weakly up and down with respect to the horizontal line. The oscillations are out of phase with respect (opposite in period) to the singlet density $\rho_w^{(1)}(z)$.  The singlet density $\rho_w^{(1)}(z)$ has a contact value $\approx 2.296$ ($z = {\sigma}/{2}$).  It oscillates vigorously with pronounced peaks and valleys.  We note that the insertion work $\beta W_{\mathrm{ins}}(z)$ also oscillates with trends similar to the pseudo-density: when $\rho_{\mathrm{pseudo}}(z)$ is up, $\beta W_{\mathrm{ins}}(z)$ is also up, and when $\rho_{\mathrm{pseudo}}(z)$ is down, $\beta W_{\mathrm{ins}}(z)$ is also down.
\begin{figure}[!h]
\centerline{\includegraphics[width=8cm]{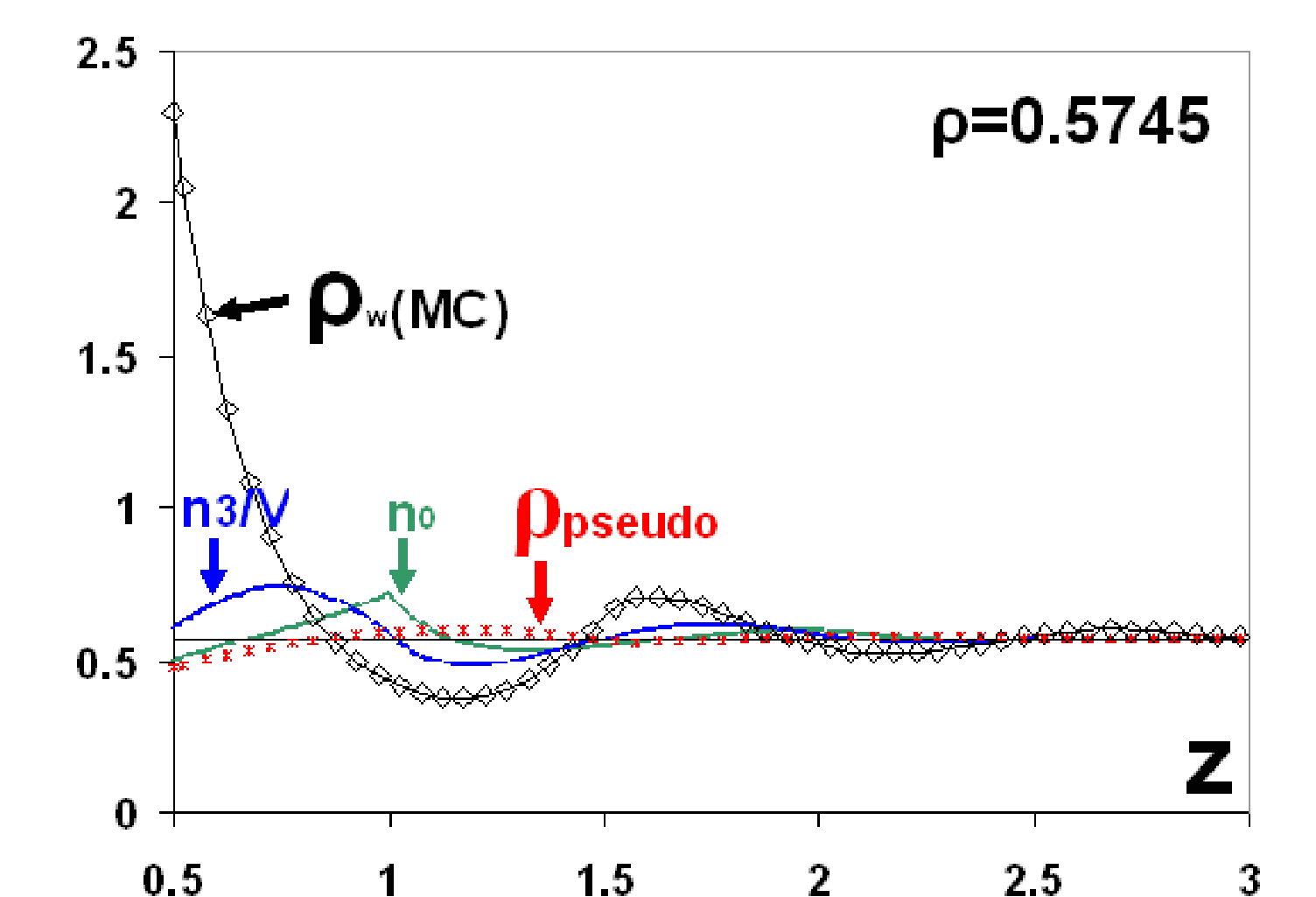}}
  \caption{Comparison of the effective (pseudo) density $\rho_{\mathrm{pseudo}}$ (symbol $\ast$) with the weighted densities $n_0$ (green line) and $n_3/V$ (blue line) from the FMT theory of Rosenfeld. Bulk hard sphere density $\rho^{\ast} = 0.5745$.  Symbol ($\diamondsuit$) $\equiv$ the singlet density profile $\rho_w^{(1)}(z)$ from MC data of Groot et al.~\protect\cite{groot}.
}
  \label{fig2}
\end{figure}

A question naturally arises: how does this pseudo-density $\rho_{\mathrm{pseudo}}(z)$ compare with the commonly used weighted densities, say, $n_{\alpha}(z)$ in the fundamental measure theory (FMT) of Rosenfeld~\cite{rosenfeld}. We selected two weighted densities $n_0(z)$ (green line) and $n_3(z)/V$ (blue line) from FMT for comparison in figure~\ref{fig2} (there are in fact six weighted densities $n_{\alpha}(z)$ in FMT ($\alpha = 0,1,2,3,V1,V2$) based on the geometries of hard spheres). The basis set of weights are composed of three functions: $\omega_2$ is based on the Dirac Delta function, thus generating $n_0$, $\omega_3$  is based on the Heaviside function, generating $n_3$, and $\omega_{V2}$ is a vector function (see Rosenfeld~\cite{rosenfeld}).  The other linearly-dependent weighting functions are related to this basis set by scaling through spherical geometries. The first three weighted densities are related by the relations: $n_0(z) = n_1(z)/R = n_2(z)/S$ ($R$ and $S$ are the radius and surface area, respectively, of a hard sphere), while $n_3(z)$ is normalized by the spherical volume $V:  n_3(z)/V$.

In figure~\ref{fig2} we notice that the magnitudes of both $n_0(z)$ and $n_3(z)/V$ are close to the pseudo-density $\rho_{\mathrm{pseudo}}(z)$ and far from the singlet density $\rho_w^{(1)}(z)$.  There are visible differences between $\rho_{\mathrm{pseudo}}(z)$ and $n_0(z)$ as well as $n_3(z)/V$.  This is not surprising, since these coarse-grained densities are derived from different theories (the FMT vs. the PDT). $\rho_{\mathrm{pseudo}}(z)$ is close to $n_0(z)$ only near the wall and at long ranges. The cusp in $n_0(z)$ (due to the delta weighting) is missing from $\rho_{\mathrm{pseudo}}(z)$.  Values of $n_3(z)/V$ are quite distinct from both $n_0(z)$ and $\rho_{\mathrm{pseudo}}(z)$. The observations also apply to other two weighted densities owing to scaling: $n_1(z)/R$ and $n_2(z)/S$. On the whole, the weighted densities and the pseudo-density are commensurate with each other (of similar magnitudes) even though they came from entirely different origins.

To make further comparisons, we also calculate the insertion work arising from the FMT.  The excess free energy $F^{\mathrm{ex}}$ in FMT according to literature~\cite{rosenfeld}) is expressed as: $\beta F^{\mathrm{ex}} = \int \rd \vec{r} \Phi(n_\alpha(\vec{r}))$  where $\Phi$ is a free energy density.  This quantity is a function of the weighted densities $n_\alpha$ and itself arises from a uniform fluid model (such as the compressibility equation from the Percus-Yevick equation~\cite{percus} (PY$^{\mathrm{c}}$) or variations thereof~\cite{kierlik}). The FMT insertion work can be evaluated as the functional derivative~\cite{kierlik} $\delta F^{\mathrm{ex}}/\delta\rho_w^{(1)}(z)$, thus in Fourier space
\begin{equation}
\label{eq26}
\beta W_{\mathrm{ins(FMT)}}(k) = \sum_{\alpha} \frac{\partial \Phi}{\partial n_\alpha}\omega_{\alpha}(k).
\end{equation}
This FMT insertion work ($W_{\mathrm{ins}}(\mathrm{FMT})$) at $\rho^{\ast} = 0.5745$ is plotted in figure~\ref{fig1} (the brown line). It quite closely follows the simulated $W_{\mathrm{ins}}(\mathrm{MC})$ (triangles).  This is to be expected, since both the FMT and the PDT-based theories are designed to match the nonuniform free energy. The FMT free energy lies somewhat above the PDT values,  because the PY$^{\mathrm{c}}$ solution for the chemical potential is higher than the MC value for hard spheres.

\begin{figure}[ht]
\centerline{\includegraphics[width=8cm]{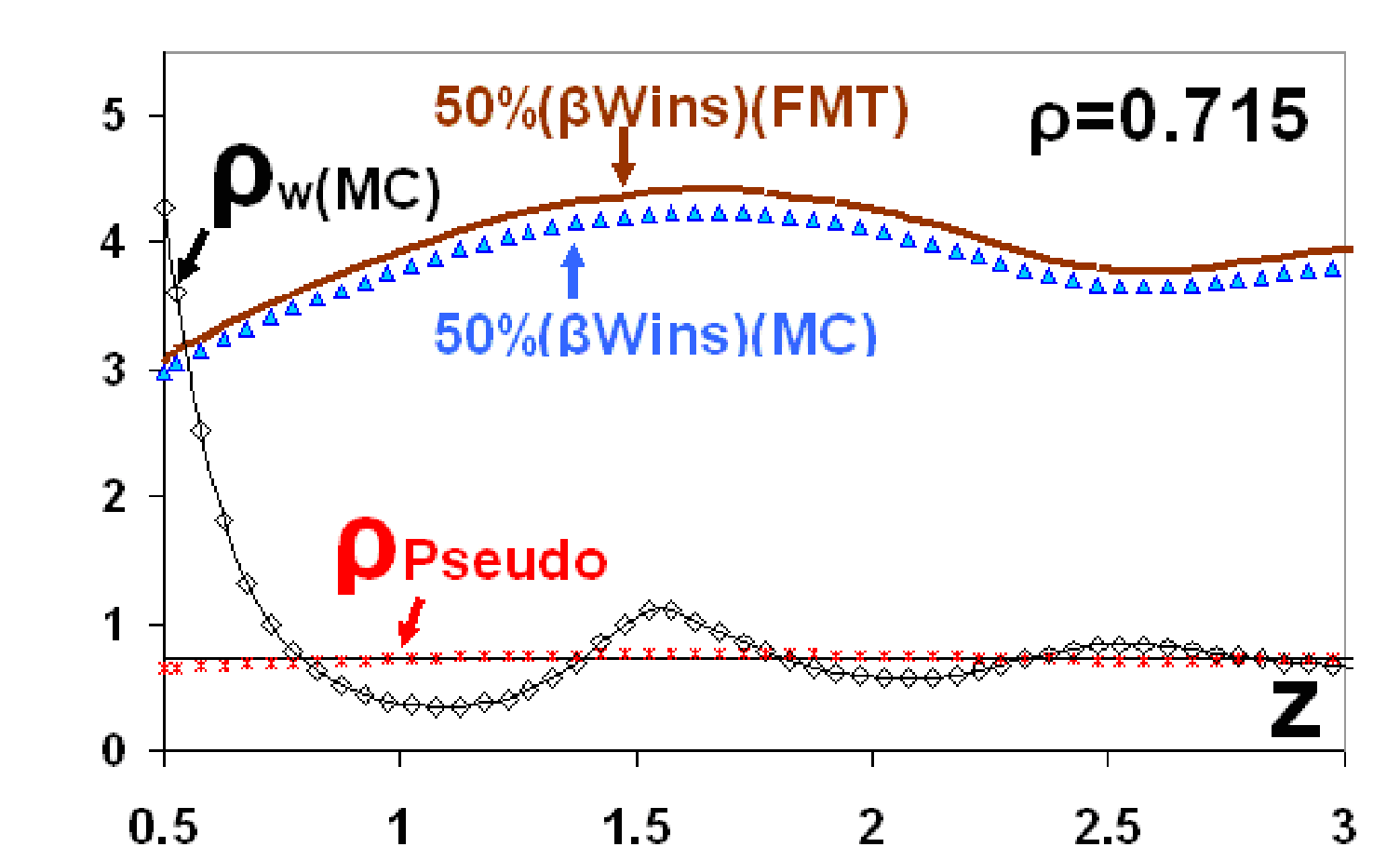}}
  \caption{The effective (pseudo) density $\rho_{\mathrm{pseudo}}$ (symbol  $\ast$; from PDT) and the insertion work $\beta W_{\mathrm{ins}}$ (symbol $\Delta$; from MC) at bulk hard sphere density $\rho^{\ast} = 0.715$ for the HS/HW system.   The insertion work is scaled down $50\%$ in order to fit in the graph.  Symbol ($\diamondsuit$) $\equiv$ the singlet density profile $\rho_w^{(1)}(z)$ from MC data of Groot et al.~\protect\cite{groot}.  The brown line is the insertion work $\beta W_{\mathrm{ins}}$ produced by the FMT theory.  The horizontal line is at $\rho^{\ast} = 0.715$. The pseudo-density $\rho_{\mathrm{pseudo}}(z)$ oscillates weakly around the bulk value and in opposite periodicity as compared to $\rho_w^{(1)}(z)$.  The insertion work $\beta W_{\mathrm{ins}}$ also oscillates synchronously with $\rho_{\mathrm{pseudo}}(z)$  but oppositely to $\rho_w^{(1)}(z)$.
}
  \label{fig3}
\end{figure}
Similar calculations are made for HS/HW at a higher density $\rho^{\ast} = 0.715$.  Figure~\ref{fig3} shows the singlet density $\rho_w^{(1)}(z)$ (diamonds). The pseudo-density (asterisks) oscillates mildly.  The FMT insertion work (brown line) is close to the MC insertion work $\beta W_{\mathrm{ins}}$ (triangles) but overestimates it.  The PY$^{\mathrm{c}}$ values are higher than the MC values as already noted.  (N.B. The values of $\beta W_{\mathrm{ins}}$ are multiplied by $0.5$ (or $50\%$), in order to fit into the same graph).

\begin{figure}[ht]
\centerline{\includegraphics[width=8cm]{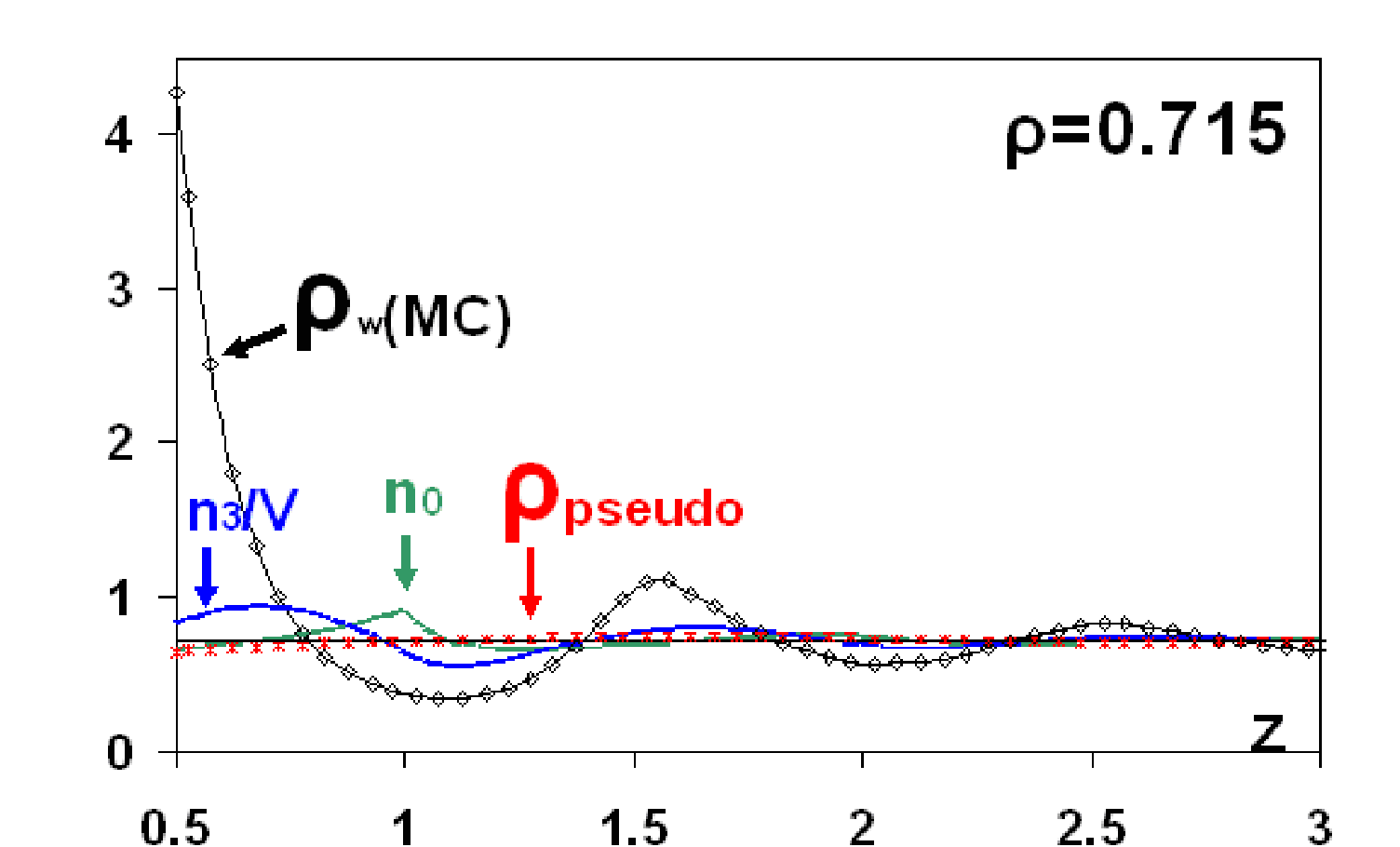}}
  \caption{Comparison of the effective (pseudo) density $\rho_{\mathrm{pseudo}}(z)$ (symbol $\ast$) with the weighted densities $n_0$ (green line) and $n_3/V$ (blue line) from the FMT theory of Rosenfeld. Bulk hard sphere density $\rho^{\ast} = 0.715$.  Symbol ($\diamondsuit$) $\equiv$ the singlet density profile $\rho_w^{(1)}(z)$ from MC data of Groot et al.~\protect\cite{groot}.
}
  \label{fig4}
\end{figure}

In figure~\ref{fig4} $\rho_{\mathrm{pseudo}}(z)$ is compared with $n_0(z)$ (green line) and $n_3(z)/V$ (blue line).  Again, we see that at a few points, $n_0(z)$ and $\rho_{\mathrm{pseudo}}(z)$ match closely, while $n_3(z)/V$ has larger differences from $\rho_{\mathrm{pseudo}}(z)$ at this higher density. The comparisons show that the PDT approach is clearly different from the fundamental measure theory.  FMT was constructed based on the PY$^{\mathrm{c}}$ theory (or modifications thereof) in order to map the compressibility free energy (see also later improved versions~\cite{roth}) to the nonuniform free energy.  In our approach, we use the PDT to map the CS free energy to the nonuniform free energy, and we use MC data to obtain the insertion work.  The objectives are the same; but the theoretical bases are different.

\begin{figure}[!h]
\centerline{\includegraphics[width=8cm]{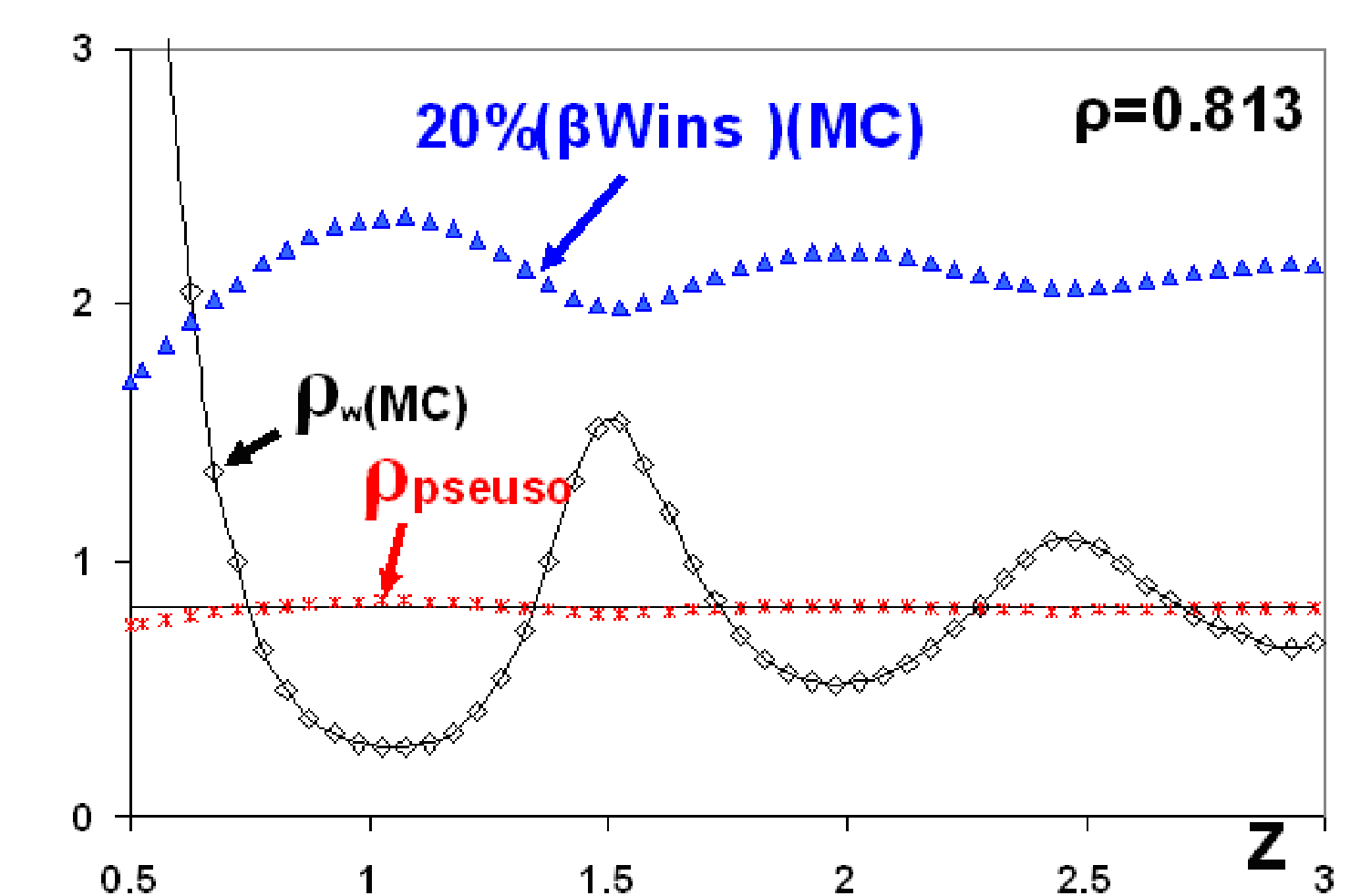}}
  \caption{The effective (pseudo) density $\rho_{\mathrm{pseudo}}$ (symbol $\ast$; from PDT) and the insertion work $\beta W_{\mathrm{ins}}$ (symbol $\Delta$; from MC)) at bulk hard sphere density $\rho^{\ast} = 0.813$ for the HS/HW system. The insertion work is scaled down to $20\%$ in order to fit in the graph.  Symbol ($\diamondsuit$) $\equiv$ the singlet density profile $\rho_w^{(1)}(z)$ from MC data of Groot et al. The horizontal line is at $\rho^{\ast} = 0.813$. The pseudo-density $\rho_{\mathrm{pseudo}}$ oscillates weakly around the bulk value and in opposite periodicity as compared to  $\rho_w^{(1)}(z)$.  The insertion work $\beta W_{\mathrm{ins}}$ also oscillates synchronously with $\rho_{\mathrm{pseudo}}(z)$ but oppositely to $\rho_w^{(1)}(z)$.
}
  \label{fig5}
\end{figure}

To further examine high-density hard spheres, we next examine the case $\rho^{\ast} = 0.813$. The results are plotted in figure~\ref{fig5}.   Similar oscillatory behavior is observed for the pseudo-density and the insertion work (reduced to $20\%$ in order to fit in the graph).  The phases of oscillations of the singlet $\rho_w^{(1)}(z)$  and $\rho_{\mathrm{pseudo}}(z)$ (and $\beta W_{\mathrm{ins}}$ as well) differ by a half period $\pi$.
\section{Construction of a closure theory}

It has been recognized that the DFT for the structures of nonuniform fluids should first be based on valid uniform-fluid properties as inputs.  In particular, the structure of nonuniform fluids depends critically on the accuracy of the pair direct correlation functions 2-DCF of the uniform fluid that must be supplied beforehand.  We have developed in the last decades an accurate theory~\cite{lee4} for the uniform hard sphere fluid: the zero-separation-theorem-based closure (ZSEP). We can thus easily generate accurate 2-DCF $C_{0\mathrm{HS}}^{(2)}$  for use in the present study of nonuniform hard-sphere fluids.

\begin{figure}[ht]
\centerline{\includegraphics[width=8cm]{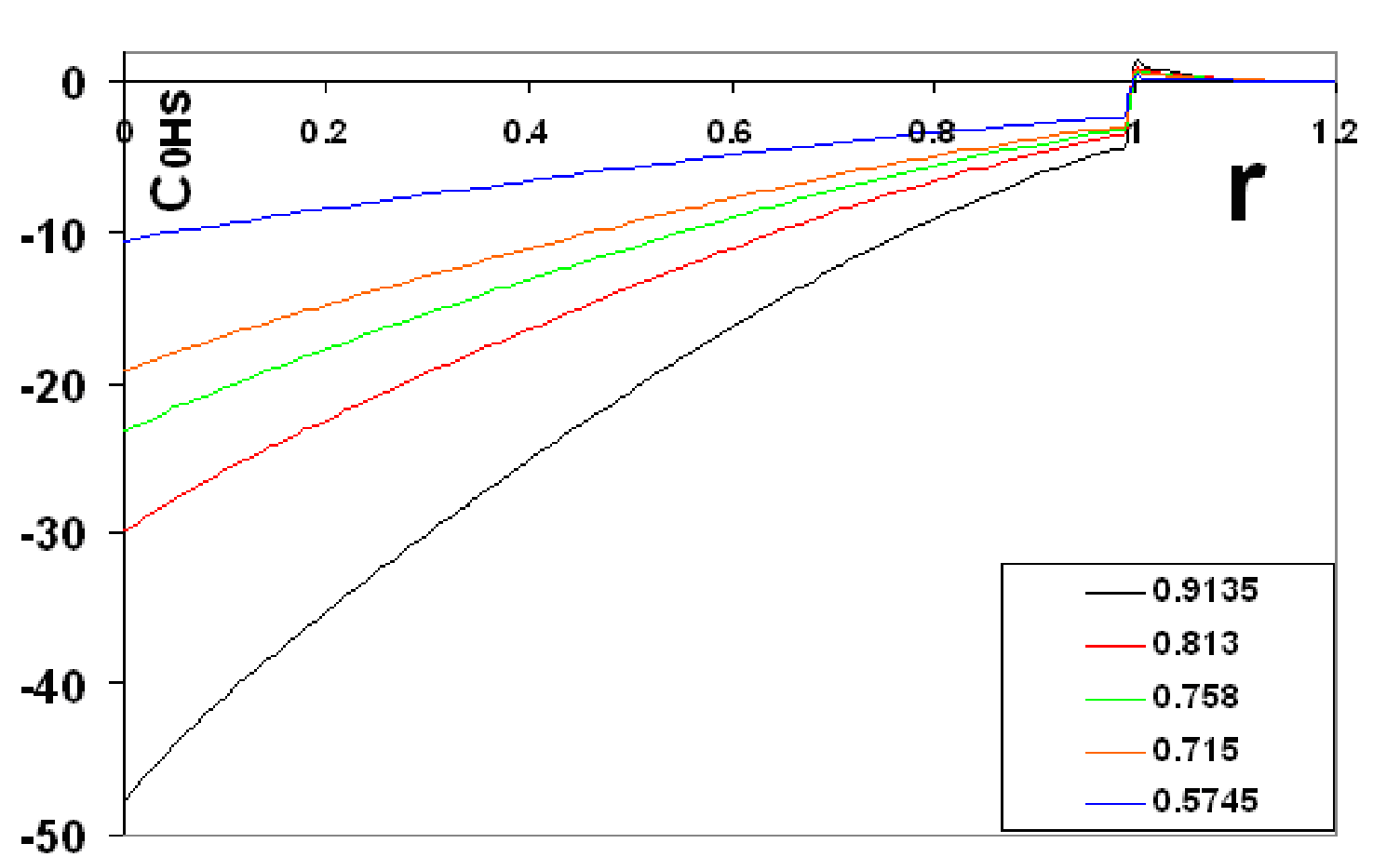}}%
  \caption{The pair direct correlation functions  $C_{0\mathrm{HS}}^{(2)}(r)$ of uniform hard spheres generated by the ZSEP closure~\protect\cite{lee4} for five densities: $\rho^{\ast} = 0.5745$, $0.715$, $0,758$, $0.813$, and $0.9135$ (Curves from top to bottom, respectively). They all show a non-zero tail at $r > 1$ (a correct behavior for hard spheres).
}
  \label{fig6}%
\end{figure}

\begin{figure}[!h]
\centerline{\includegraphics[width=8cm]{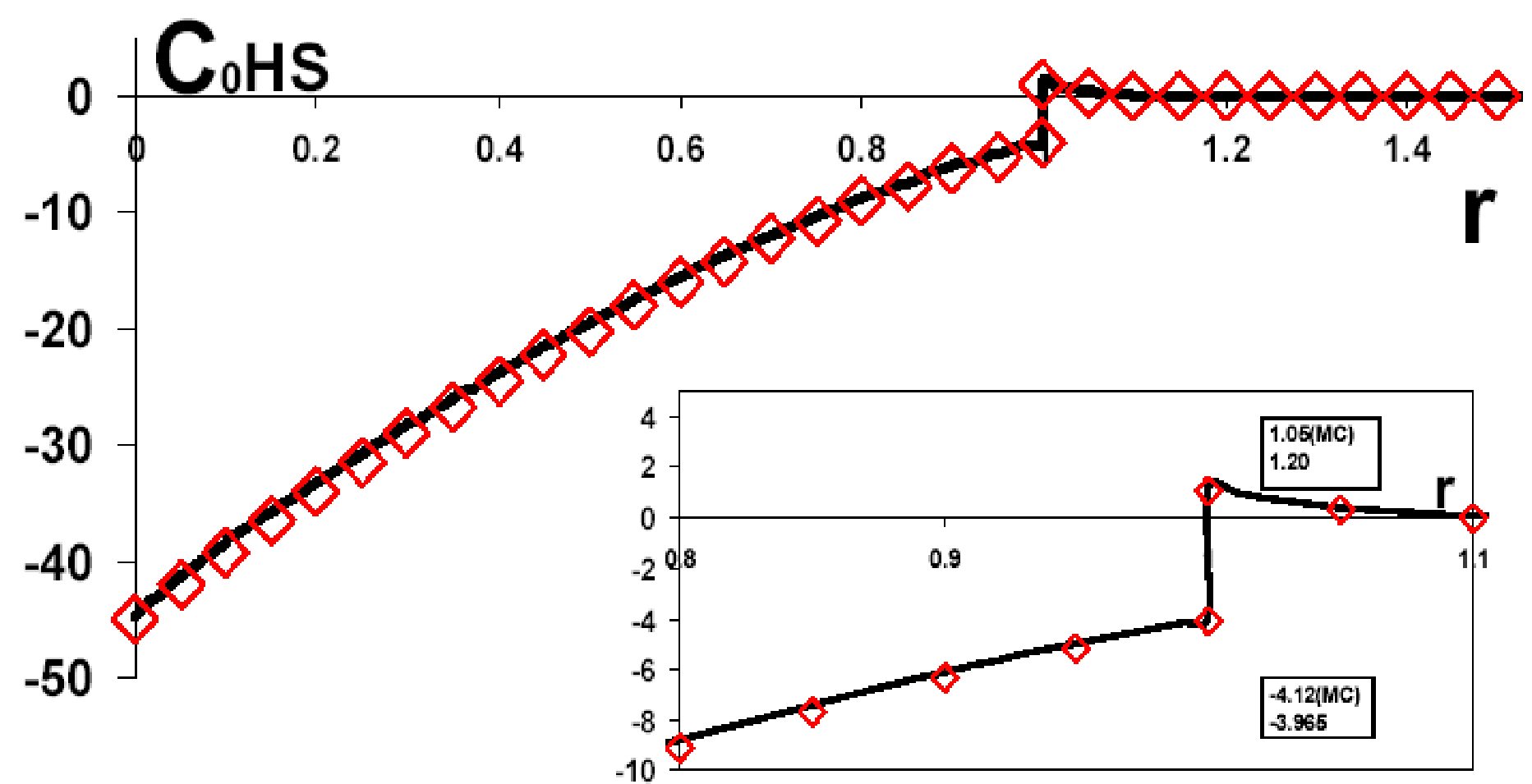}}
  \caption{The pair direct correlation functions $C_{0\mathrm{HS}}^{(2)}(r)$ of uniform hard spheres generated by the ZSEP closure~\protect\cite{lee4} for the bulk density $\rho^{\ast} = 0.90$.  The lines are from ZSEP theory, and the symbols ($\diamondsuit$) from MC data of Groot~\protect\cite{groot2}.  The inset shows the magnified view of the non-zero tail and discontinuity in $C_{0\mathrm{HS}}^{(2)}(r)$.  At $r = \sigma^+$,  $C_{0\mathrm{HS}}^{(2)}(\sigma^+) = 1.20$ (from ZSEP), and $1.05$ (from MC).  At $r = \sigma^-$, $C_{0\mathrm{HS}}^{(2)}(\sigma^-) = -3.965$ (from ZSEP), and $-4.12$ (from MC).
}
  \label{fig7}%
\end{figure}

We shall look at the five densities where the MC data are available for the HS/HW systems: namely $\rho^{\ast} = 0.5745$, $0.715$, $0.758$, $0.813$, and $0.9135$.  The ZSEP closure for the uniform bridge function $B_{0\mathrm{HS}}$ is of the form~\cite{lee4} (caret indicates a function)
\begin{equation}
\label{eq27}
\hat{B}_{0\mathrm{HS}}(\gamma^{\ast}) = - \frac{\zeta}{2}\gamma^{\ast2}\left[ 1 - \phi + \frac{\phi}{1 + \alpha \gamma^{\ast}}\right],
\end{equation}
where $\alpha$, $\phi$, and $\zeta$ are adjustable parameters, and $\gamma^{\ast} = \gamma + \rho \frac{f}{2}$.  $f$ is the Mayer factor of the repulsive WCA (Weeks-Chandler-Andersen) potential. For details, see references in~\cite{lee4}. The adjustable parameters $\alpha$, $\phi$, and $\zeta$ are determined using structural consistencies (the zero-separation theorem and the contact-value theorem) and thermodynamic consistencies (e.g., the pressure consistency, and Gibbs-Duhem relation).  The uniform 2-DCF's $C_{0\mathrm{HS}}^{(2)}$ obtained are depicted in figure~\ref{fig6}. They have all satisfied closely the consistency relations imposed.  For example, in the case of the high density $\rho^{\ast} = 0.90$, the $C_{0\mathrm{HS}}^{(2)}$ is compared with the MC-generated data~\cite{groot2} in figure~\ref{fig7}.
Diamond symbols denote the MC data, and the solid line is the 2-DCF from the ZSEP(\ref{eq27}).  We see very close agreement of the two curves.  The zero-value $C_{0\mathrm{HS}}^{(2)}$  of MC is $-44.90$ while the ZSEP value is $-47.70$ (within $6\%$).  In particular, the values $C_{0\mathrm{HS}}^{(2)}(r)$ derived from ZSEP for $r > 1$ are positive and non-zero.  This is corroborated by the MC data.  The positive peak  $C_{0\mathrm{HS}}^{(2)}(\sigma^+) = 1.2$ from ZSEP, while MC gives $1.05$.  This is in obvious contrast to the PY-based theories that would produce $C_{0\mathrm{HS}}^{(2)}(r) = 0$ for $r > 1$ (an erroneous result).  The same observation applies to all five present cases (figure~\ref{fig6}).

For the nonuniform HS/HW case, we developed a closure similar to ZSEP~\cite{lee2}:
\begin{equation}
\label{eq28}
\hat{B}_{w}(\gamma_w) = - \mathrm{sgn}(\gamma_w) \cdot \zeta' \gamma_w^{2}\left[ 1 - \phi' + \frac{\phi'}{1 + \alpha' \gamma_w}\right],
\end{equation}
where $\alpha'$, $\phi'$, and $\zeta'$ are adjustable parameters for the nonuniform $B_w$.  $\mathrm{sgn}(.)$ is the sign function. $\gamma_w$ is the indirect correlation function defined as
\begin{equation}
\label{eq29}
\gamma_w(\vec{r}) \equiv \int \rd \vec{r}\,' C_{0\mathrm{HS}}^{(2)}(\lvert \vec{r} - \vec{r}\,' \rvert) \delta \rho_w(\vec{r}\,').
\end{equation}

We found from practice that ``renormalization'' of the indirect correlation is not needed for this nonuniform fluid.  We also note that the expression~(\ref{eq28}) is modeled after the ZSEP equation~(\ref{eq27}) for uniform fluids~\cite{lee4}.  In earlier studies~\cite{lee1,lee2} we have shown that this closure was effective for adsorption of high-density Lennard-Jones fluids. The similarity between the uniform closure and the nonuniform closure used represents an ongoing effort in the transfer of successful theories for uniform liquids to nonuniform liquids.

\begin{table}[ht]
\caption{Parameters for the ZSEP closure \eqref{eq28} and wall sum rule for the HS/HW system. }
\begin{center}
    \begin{tabular}{lccccc} \\
        \cline{1-6}
       $\rho_{0}$ &  $0.5745$ & $0.715$ &
             $0.758$ & $0.813$   & $0.9135 $     \\
        \hline
    $\alpha'$   & $0.4$ & $0.4$ & $0.4$ &
        $0.5$ & $0.62$   \\
    $\phi'$   & $0.4$ & $0.4$ & $0.4$ &
        $0.4$ & $0.4$   \\
    $\zeta'$   & $-0.38$ & $-0.28$ & $-0.22$ &
        $-0.20$ & $-0.13$   \\
        \hline
  $^\dag\rho_w^{(1)}({\sigma}/{2})$ & 2.288    & 4.319       & 5.131       &  6.598      & 10.23   \\
  \hline
  $^\S\rho_w^{(1)}({\sigma}/{2})$   & 2.29     & 4.27        & 5.15        &  6.57       & 10.28   \\
        \hline
    \end{tabular}
  \label{tab2}
  \end{center}
  \footnotesize{$^\dag$Contact value (from ZSEP calculations). \\$^\S$Contact value (from hard-wall sum rule: $\rho_w^{(1)}({\sigma}/{2}) = P/kT$) from Carnahan-Starling equation.}
\end{table}
Equations~(\ref{eq6}),~(\ref{eq28}), and~(\ref{eq29}) are coupled. The convolution was solved by standard numerical methods~\cite{lee2} in bipolar coordinates.  Trapezoidal rule was used in  integration. Picard's iterations with mixing (relaxation) on input and output iterates were employed.  The grid of numerical integration was $\Delta r = 0.01 \sigma$, and total grid number $N=2048$.  Thus, the maximum $z$-distance reached is $20.48\sigma$.  Cauchy's absolute convergence was enforced for the $\gamma_w$-function with convergence criterion $\delta = 0.0001$.  There were three free parameters to be determined in equation~(\ref{eq28}): $\alpha'$, $\phi'$, and $\zeta'$.  We needed three conditions.   The first condition was the hard-wall sum rule: the contact density $\rho_w^{(1)}({\sigma}/{2}) = P/(kT)$.  In practice, we set $\phi'$ to a default value of $0.4$.  $\alpha'$ and $\zeta'$ were adjusted in tandem until the wall rule was satisfied (see table~\ref{tab2}). In the future, other sum rules~\cite{henderson2_2} will be tested and incorporated.

\begin{figure}[ht]
\centerline{\includegraphics[width=8cm]{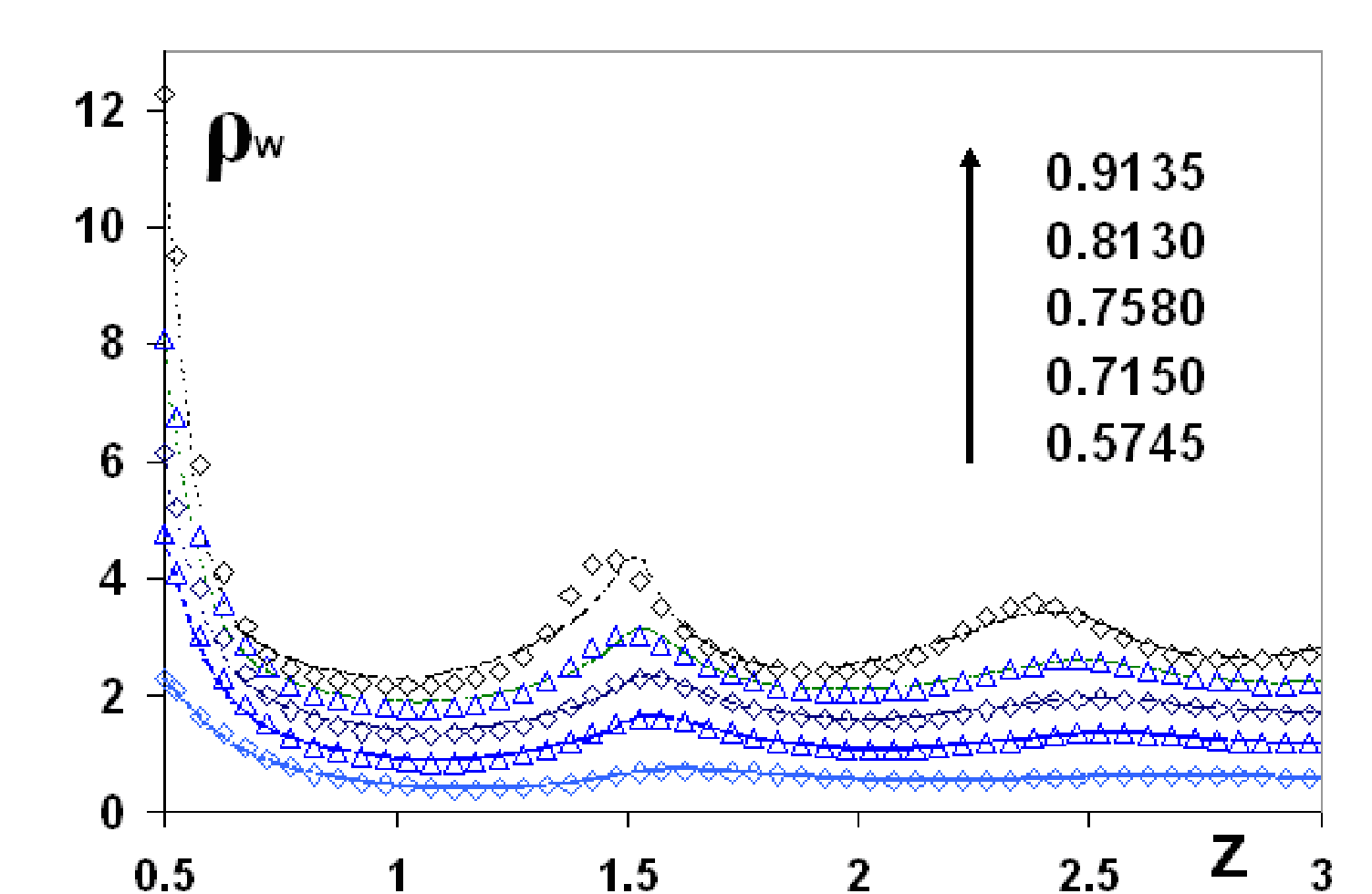}}
  \caption{The nonuniform singlet density functions $\rho_w^{(1)}(z)$ obtained from the new ZSEP-closure equation~(~\protect\ref{eq28}) for the HS/HW system at five densities $\rho^{\ast} = 0.5745$, $0.715$, $0.758$, $0.813$, and $0.9135$ (from bottom to top. Each curve shifted up by 0.5 units, respectively). Symbols = MC data from Groot et al. [15].  Lines $\equiv$ ZSEP results. We discern close agreement between the ZSEP curves and the MC curves except at the highest density ($0.9135$) where the second peak of the theoretical curve is displaced to a larger z-value. (See discussions in the text).
}
  \label{fig8}
\end{figure}
\begin{figure}[!h]
\centerline{\includegraphics[width=8cm]{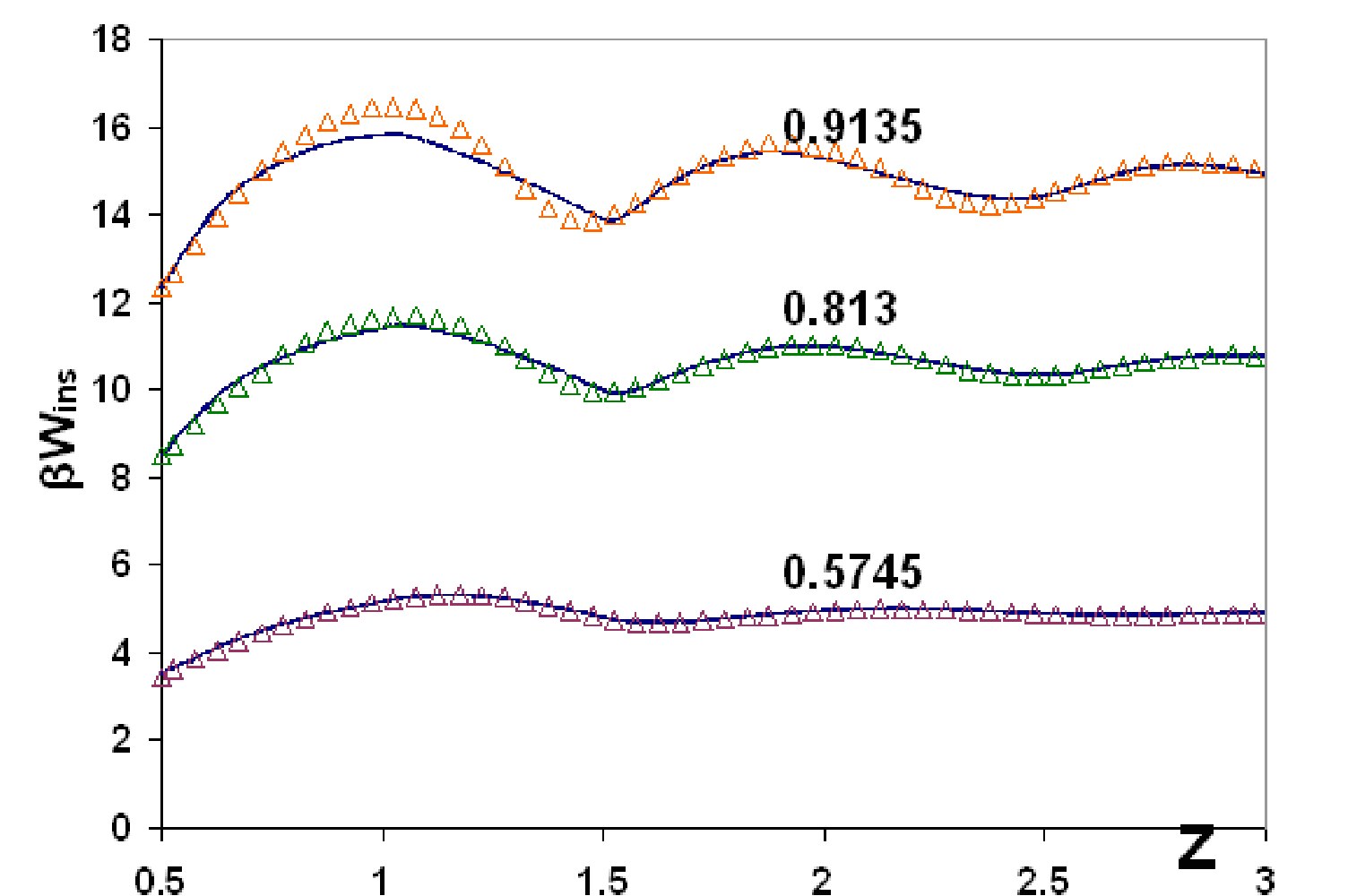}}
  \caption{The insertion work generated by the new closure theory at $\rho^{\ast} = 0.5745$, $0.813$, and $0.9135$ (Curves from bottom to top, respectively).  Lines $\equiv$ ZSEP theory; triangles $\equiv$ MC data. Close agreement is in evidence for $\rho^{\ast} = 0.5745$ and $0.813$.  The deviations at $\rho^{\ast} = 0.9135$ reflect the same cause: the two-dimensional solid-fluid transition of the surface layer of hard spheres at this high density.
}
  \label{fig9}
\end{figure}
Figure~\ref{fig8} shows the result of application of ZSEP.  The density varies from the moderate $\rho^{\ast} = 0.5745$ to the highest $\rho^{\ast} = 0.9135$. The symbols are the MC data~\cite{groot}, the lines are from the ZSEP theory.  We observe overall superb agreement for densities $\rho^{\ast}$ from $0.5745$, $0.715$, $0.758$ to $0.813$.  At $\rho^{\ast} = 0.813$, our results compare well with previous accurate theoretical attempts (e.g. the White-Bear version of FMT~\cite{roth}).  The ingredients for success are from (i) the high accuracy of the input uniform 2-DCF's; and (ii) the effectiveness of the ZSEP closure~(\ref{eq28}).  For the highest density $\rho^{\ast} =  0.9135$ there is some deterioration of the theoretical curve at the second peak. The heights of the peaks (of ZSEP and MC) are similar, however the predicted location is displaced to a longer distance. A closer reading of the original MC source~\cite{groot} reveals that at this high density (packing fraction $\eta = 0.4783$), there are clusters of molecules forming hexagonal structures for the first layer of particles (figure~\ref{fig5} of \cite{groot}), and reduced normal-direction diffusion coefficient for the surface molecules: all indications of two-dimensional quasi-solid-fluid transition.  Note that the fluid limit of homogenous hard spheres is at about $\eta \approx 0.4948$. For inhomogeneous hard spheres this may occur earlier.  This puts stringent demand on a closure theory which is not intended to predict solid formation.

Next, we calculate the insertion work via equation~(\ref{eq22}) using ZSEP-density $\rho_w^{(1)}(z)$ as input.  The results are shown in figure~\ref{fig9} for three densities  $\rho^{\ast} = 0.5745$, $0.813$, and $0.9135$.  The agreement between the ZSEP $\beta W_{\mathrm{ins}}$ (lines) and MC $\beta W_{\mathrm{ins}}$ (triangles) is excellent for $\rho^{\ast} = 0.5745$ and $0.813$.  For $\rho^{\ast} = 0.9135$, similar discrepancies as for the singlet density $\rho_w^{(1)}(z)$ are in evidence.  The displacement of the first minimum to a larger distance corresponds to the displacement of the second peak of $\rho_w^{(1)}(z)$.

\section{Conclusions}

The two major questions that we set out to explore in this study are (i) how does the effective density based on the potential distribution theorem look like and how does it behave? (ii) Can we obtain the structure of hard spheres adsorbed on the hard wall through a closure theory for nonuniform fluids?  We formulate two potential distribution theorems: one for the insertion work of the usual test particle that interacts with the wall, and the other for the test particle that \emph{ignores} the wall.  It is the latter insertion work that corresponds to the singlet direct correlation and as a consequence the second potential distribution theorem becomes equivalent to the Euler-Lagrange equation.

The pseudo-densities $\rho_{\mathrm{pseudo}}$\,, due to their construction, are much smoother quantities than the singlet densities.  They oscillate around the bulk value with periods out-of phase with respect to the nonuniform $\rho_w^{(1)}$.

The insertion work is derived from the MC data on HS/HW (Groot et al., 1987~\cite{groot}) via the Euler-Lagrange equation (with the identification $\beta W_{\mathrm{ins}} = - C_{w}^{(1)}$ ). Its behavior for several densities ($\rho^{\ast} = 0.5745$, $0.715$, and $0.813$) is determined. $\beta W_{\mathrm{ins}}$ oscillates in phase as the pseudo-density, but out of phase with respect to the singlet density $\rho_w^{(1)}$.

The density functional theories need accurate uniform pair direct correlation functions $C_{0\mathrm{HS}}^{(2)}$ as input.  Earlier we have formulated an accurate closure theory for the 2-DCF (the ZSEP theory) of uniform hard spheres.  It is computationally cheap now to generate these functions: it took less than a minute of CPU time on a PC computer to produce an accurate DCF.  Here we propose a new closure theory for the nonuniform hard spheres equation~(\ref{eq28}).  Its form is inspired by the uniform fluid ZSEP equation. This bridge function is used in the star-function based density functional theory equation~(\ref{eq6}) to generate the singlet density profiles.  These ZSEP densities are compared with the MC data (figure~\ref{fig8}) at five densities $\rho^{\ast} = 0.5745$, $0.715$, $0,758$, $0.813$, and $0.9135$.  Except for the highest density, the ZSEP gives accurate nonuniform density profiles.  This shows that the closure-based density functional theory can perform reasonably well for hard spheres when a suitable closure (bridge function) has been adopted.  We shall have further developments on this in the future.

These theoretical predictions for the singlet densities are used to calculate the insertion works (figure~\ref{fig9}).  They also compare favorably with the MC data.

In summary, we have proposed and evaluated a new type of effective densities via the potential distribution theorem. We characterize their behavior for the hard spheres/hard wall system at moderate to high densities (up to $\rho^{\ast} = 0.9135$).  The free energies (the insertion works $W_{\mathrm{ins}}$) of the nonuniform system are also calculated.  They show oscillations opposite in phase to those of the singlet densities. We also construct a new closure-based density functional theory that can give accurate reproduction of the computer simulated densities and insertion works.

The newly developed star-based density functional theory can be extended to fluids interacting with soft/attractive potentials. This task is under active investigation.

\section*{Acknowledgements}
G.P. acknowledges the Competitive Grant support by the University of Kwazulu-Natal.
L.L.L. had stimulating discussions with Walter Chapman and Ken Cox during my sabbatical leave at Rice University in 2011.  He also had useful discussions with Zhengzheng Feng on the White-Bear calculations of the FMT theory.  L.L.L. thanks them for their hospitality.

\newpage

\ukrainianpart

\title{Адсорбція твердих сфер: структура та ефективна густина відповідно
до теореми розподілу потенціалу
}
\author{Л.Л. Лі\refaddr{label1}, Г.~Пелікане\refaddr{label2}}
\addresses{
\addr{label1} Відділ хімічної інженерії та інженерії матеріалів,
Університет Каліфорнії, \\Помона, Каліфорнія, США
\addr{label2} Школа фізики, Університет імені Квазулу-Наталь, Скотсвіль, \\
3209 Пітермаріцбург, Південна Африка
}
\makeukrtitle

\begin{abstract}
Ми пропонуємо новий тип ефективних густин, отриманих з теореми розподілу
потенціалу. Ці гус\-ти\-ни необхідні для відображення вільної енергії
\emph{однорідної} рідини у вільну енергію \emph{неоднорідної} рідини.
Теорема розподілу потенціалу дає роботу, необхідну для устромляння
 пробної частинки у систему молекул, на яку діє зовнішній потенціал. Ця
робота $W_{\mathrm{ins}}$ може бути отримана з симуляції Монте-Карло
(MК) (наприклад, підходом тестової частинки Відома) або з аналітичної
теорії. Псевдогустини є побудовані так, що коли їх значення
підставляються в рівняння стану \emph{однорідної} рідини (наприклад,
рівняння для  хімічних потенціалів Карнагана-Старлінгa системи твердих сфер),
то відтворюється робота устромляння частинки в неоднорідну рідину, отримана з
симуляції МК. Ми досліджуємо поведінку псевдогустини для системи ``тверді сфери''-``тверда стінка'' при середніх та високих густинах (від $\rho^{\ast} =
0.5745$ до $0.9135$). Для цього використовуються результати Монте-Карло
Гроота та співавторів. Псевдогустини демонструють осцилюючу поведінку з протилежною фазою до
одночастинкових густин. Ми також пропонуємо нову теорію функціоналу густини на
основі замикання (\emph{теорію функціоналу густини зіркової функції}),
що може точно описувати профілі густини і роботу устромляння. Точність теорії
перевіряється для декількох випадків адсорбції твердих сфер.
\keywords теорема розподілу потенціалу, тверді сфери, адсорбція, ефективна
густина, структура, інтегральне рівняння, замикання
\end{abstract}

\end{document}